\documentclass[useAMS,usenatbib,onecolumn,usegraphicx]{mn2e}

\newcommand{\beq}{\begin{equation}}
\newcommand{\eeq}{\end{equation}}
\newcommand{\beqar}{\begin{eqnarray}}
\newcommand{\eeqar}{\end{eqnarray}}

\title[Torsional Oscillations of Relativistic Stars II]
{Torsional Oscillations of Relativistic Stars with Dipole Magnetic
  Fields II. Global Alfv\'en Modes}
\author[H.~Sotani, K.~D. Kokkotas, N.~Stergioulas, and M.~Vavoulidis]
{H.~Sotani\thanks{E-mail:
sotani@astro.auth.gr}, K.~D.~Kokkotas\thanks{E-mail:
kokkotas@auth.gr},  N.~Stergioulas\thanks{E-mail:
niksterg@astro.auth.gr},
and M. Vavoulidis\thanks{E-mail:
miltos@astro.auth.gr}
\\
  Department of Physics, Aristotle University of Thessaloniki,
  Thessaloniki 54124, Greece}
\begin{document}

\maketitle

\label{firstpage}

%%%%%%%%%%%%%%%%%%%%%%%%%%%%%%%%%%%%%%%%%%%%%%%%%%%%%%%%%%%%%%%%%%%%%%%%%%%%
% Abstract
\begin{abstract}
  We investigate torsional Alfv\'{e}n modes of relativistic stars with
  a global dipole magnetic field. It has been noted recently
  (Glampedakis et al.  2006) that such oscillation modes could serve
  as as an alternative explanation (in contrast to torsional
  crustal modes) for the SGR phenomenon, if the magnetic field is not
  confined to the crust. We compute global Alfv\'{e}n modes for a
  representative sample of equations of state and magnetar masses, in
  the ideal MHD approximation and ignoring $\ell \pm 2$ terms in the
  eigenfunction. We find that the presence of a realistic crust has a
  negligible effect on Alfv\'{e}n modes for $B > 4\times 10^{15}$ G.
  Furthermore, we find strong avoided crossings between torsional
  Alfv\'{e}n modes and torsional crust modes. For magnetar-like
  magnetic field strengths, the spacing between consecutive Alfv\'{e}n
  modes is of the same order as the gap of avoided crossings. As a
  result, it is not possible to identify modes of predominantly crustal
  character and all oscillations are predominantly Alfv\'{e}n-like.
  Interestingly, we find excellent agreement between our computed
  frequencies and observed frequencies in two SGRs, for a maximum magnetic
  field strength in the range of (0.8--1.2)$\times 10^{16}$ G.
\end{abstract}
%%%%%%%%%%%%%%%%%%%%%%%%%%%%%%%%%%%%%%%%%%%%%%%%%%%%%%%%%%%%%%%%%%%%%%%%%%%%


%%%%%%%%%%%%%%%%%%%%%%%%%%%%%%%%%%%%%%%%%%%%%%%%%%%%%%%%%%%%%%%%%%%%%%%%%%%%
%%%%%%%%%%%%%%%%%%%%%%%%%%%%%%%%%%%%%%%%%%%%%%%%%%%%%%%%%%%%%%%%%%%%%%%%%%%%
\section{Introduction}
\label{sec:Intro}
%%%%%%%%%%%%%%%%%%%%%%%%%%%%%%%%%%%%%%%%%%%%%%%%%%%%%%%%%%%%%%%%%%%%%%%%%%%%
%%%%%%%%%%%%%%%%%%%%%%%%%%%%%%%%%%%%%%%%%%%%%%%%%%%%%%%%%%%%%%%%%%%%%%%%%%%%

Soft Gamma Repeaters (SGRs) have attracted a lot of attention
recently, since there exist at least two sources in which
quasi-periodic oscillations have been observed. The frequency of these
oscillations, in the tail of the burst, indicate that they must
originate in the vicinity of the SGR source, thought to be a
magnetar \citep{DT1992}. A popular model for explaining these
oscillations involves the excitation of torsional modes in the solid
crust of the star \citep{Duncan1998}. An alternative explanation has
been provided by \citet{GSA2006}. In a simple toy model, they showed
that, in addition to crustal modes, a magnetized star could also have
a discrete spectrum of global Alfv\'{e}n modes, which would dominate
over crustal modes for sufficiently strong magnetic fields. Here we
investigate global, torsional Alfv\'{e}n modes in realistic models of
relativistic stars with a global dipole magnetic field and compare our
computed frequencies with observations.

SGRs produce giant flares with peak luminosities of $10^{44}$ --
$10^{46}$ erg/s, which display a decaying tail for several hundred
seconds.  Up to now, three giant flares have been detected, SGR 0526-66
in 1979, SGR 1900+14 in 1998, and SGR 1086-20 in 2004.  The timing
analysis of the latter two events revealed several quasi-periodic
oscillations (QPOs) in the decaying tail, with frequencies of 18, 26,
29, 92.5, 150, 626.5, and 1837 Hz for SGR 1806-20 and 28, 54, 84, and
155 Hz for SGR 1900+14 (see \cite{SW2006} and references therein).
Furthermore, for SGR 1806-20 the possible detection of two additional
higher frequencies (720 Hz and 2384 Hz) has been reported
 \citep{SW2006}.

 Spherical stars have generally two type oscillations, spheroidal
 modes with even parity and the toroidal modes with axial parity. The
 observed QPOs in SGR tails are thought to originate from toroidal
 oscillations, since the latter are excited more easily than poloidal
 oscillations, as toroidal modes do not involve density variations. In
 Newtonian theory, there have been several investigations of torsional
 modes in neutron star crusts, see e.g.
 \cite{HC1980,McDermott1988,Carroll1986,Strohmayer1991a,Lee2006}.  On
 the other hand, only a few studies have taken general relativity into
 account \cite{Schumaker1983,Leins1994,Messios2001,Sotani2006,SA2006}.
 The more recent of these studies (see e.g. \cite{Sotani2006},
 hereafter Paper~I) have shown that some of the observational data of
 SGRs could agree with the crust torsional modes, if, e.g.,
 frequencies lower than 155 Hz are identified with the fundamental
 modes of different harmonic index $\ell$, while higher frequencies
 are identified with overtones.  However, as mentioned in Paper~I it
 will be quite challenging to identify all observed QPO frequencies
 with crustal torsional modes.  For example, it is difficult to
 explain both the frequencies of 26 and 29 Hz for SGR 1806-20 for a
 single stellar model, because the actual spacing of torsional modes
 of the crust is larger than the difference between these two
 frequencies. Similarly, the spacing between the 626.5 and 720 Hz QPOs
 in SGR 1806-20 may be too small to be explained by consecutive
 overtones of crustal torsional modes. In the present paper, we show
that such difficulties are avoided if one identifies the observed
frequencies with global Alfv\'{e}n modes.

 The article is structured as follows: in the next section \ref{sec:II}
we briefly summarize the equations to be solved and the assumptions made. 
In section \ref{sec:III} we discuss the numerical
results, where to understand qualitatively the properties of global
Alfv\'{e}n modes we also treat two special cases. The comparison
with the observed data in SGRs are done in section \ref{sec:IV} and
finally we give a summary and discussion in section \ref{sec:V}.
Unless otherwise noted, we adopt units of $c=G=1$,
 where $c$ and $G$ denote the speed of light and the gravitational
 constant, respectively, while the metric signature is $(-,+,+,+)$.

%%%%%%%%%%%%%%%%%%%%%%%%%%%%%%%%%%%%%%%%%%%%%%%%%%%%%%%%%%%%%%%%%%%%%%%%%%%%
%%%%%%%%%%%%%%%%%%%%%%%%%%%%%%%%%%%%%%%%%%%%%%%%%%%%%%%%%%%%%%%%%%%%%%%%%%%%
\section{Main Equations and Numerical Method}
\label{sec:II}
%%%%%%%%%%%%%%%%%%%%%%%%%%%%%%%%%%%%%%%%%%%%%%%%%%%%%%%%%%%%%%%%%%%%%%%%%%%%
%%%%%%%%%%%%%%%%%%%%%%%%%%%%%%%%%%%%%%%%%%%%%%%%%%%%%%%%%%%%%%%%%%%%%%%%%%%%

The main equations and numerical method required for the present
study have already been derived in Paper I. Here, we only give a
brief summary of the main assumptions and final relations.

We assume that the equilibrium stellar model is described by a
solution of the TOV equations for nonrotating, relativistic stars
and a metric of the form
\begin{equation}
 ds^2 = -e^{2\Phi}dt^2 + e^{2\Lambda}dr^2 + r^2(d \theta^2 + \sin^2\theta
d \phi^2), \label{metric}
\end{equation}
where $\Phi$ and $\Lambda$ are functions of the Schwarzschild
radial coordinate $r$. We neglect the influence of the magnetic field
on the equilibrium configuration, since the magnetic energy ${\cal
  E_M}$ is negligible, compared to the gravitational binding energy
${\cal E_G}$, ${\cal E_M}/{\cal E_G} \approx 10^{-4}
(B/(10^{16}G))^2$.

Adopting the ideal MHD approximation, we assume that the star is
endowed with a dipolar magnetic field, described by
\begin{eqnarray}
 H_r &=&
     \frac{e^{\Lambda}\cos\theta}{\sqrt{\pi}r^2}a_1\, , \label{eq:H_r} \\
 H_{\theta} &=&
    - \frac{e^{-\Lambda}\sin\theta}{\sqrt{4\pi}} {a_1}_{,r}.
\label{eq:H_theta}
\end{eqnarray}
where $H_\mu \equiv B_\mu/\sqrt{4\pi}$, with $B_\mu$ the magnetic
field 4-vector, and $a_1$ being a function describing the radial
behavior of the magnetic field. In the exterior of the star
\begin{equation}
 a_1^{\rm (ex)} = -\frac{3\mu_b}{8M^3}r^2 \left[\ln\left(1-\frac{2M}{r}\right)
 + \frac{2M}{r} + \frac{2M^2}{r^2}\right],
\end{equation}
where $\mu_b$ is the magnetic dipole moment for an observer at infinity
\citep{Wasserman1983}, while the interior solution is found by solving
Maxwell's equations for the $\phi$-component of the electromagnetic
4-potential $A_\mu$, assuming a particular form for the 4-current $J_\phi$
\citep{Konno1999}.

We have used a variety of neutron star models, using four different
equations of state for the core, ranging from a very soft EoS (EoS
A) \citep{EOS_A} to a very stiff (EoS L) \citep{EOS_L}, with two
intermediate ones, EoS WFF3 \citep{EOS_WFF3} and APR
\citep{EOS_APR}. For each EoS we constructed a number of models,
starting from a gravitational mass of $1.4 M_\odot$ and reaching
close to the maximum mass limit in increments of $0.2 M_ \odot$. In
order to separately investigate the effect of the composition of the
crust, we matched the various high-density EoS to two different
proposed equations of state for the crust, one recent derived by
\citet{DH2001} (DH) and, for reference, and older EoS by
\citet{NV1973} (NV). The two crust EoS differ significantly both in
the detailed composition, as well as in the density at the base of
the crust, which is at $\rho\approx 2.4\times 10^{14}$gr/cm$^3$ for
\citet{NV1973} and at $\rho\approx 1.28\times 10^{14}$gr/cm$^3$ for
\citet{DH2001}. For the stiff EoS these different properties of the
crust have a considerable effect on the bulk properties. Figure 1
%\ref{Fig:EoS}
of Paper~I displays the mass-radius relationship of
nonrotating equilibrium models constructed with the various
high-density EoS, in combination with our two choices for the crust
EoS. The individual models for which we compute torsional modes are
shown with symbols, while their detailed properties are also listed
in Table 1 of Paper~I.

For studying individual oscillation modes, the linearized dynamical
equations governing the magnetized fluid and crust are specialized
to the case of axial perturbations, adopting the relativistic
Cowling approximation (i.e. neglecting spacetime perturbations). In
this case, the only nonvanishing perturbed fluid variable is
(assuming a harmonic time dependence with frequency $\omega$)
\begin{equation}
 \delta u^{\phi} = i \omega e^{-\Phi} {\cal Y}(r) e^{i\omega t} b(\theta),
\end{equation}
where
\begin{equation}
 b(\theta) \equiv \frac{1}{\sin\theta}\partial_{\theta}
P_{\ell}(\cos\theta)\, ,
\end{equation}
and $\partial_\theta$ denotes the partial derivative with
respect to $\theta$.
Above, ${\cal Y}(r)$ describes the radial dependence of the angular
displacement of the stellar material, $P_{\ell}(\cos\theta)$ is the
Legendre polynomial of order $\ell$ and we have set $m=0$, due to
the degeneracy in $m$ for a spherically symmetric background.

%\begin{figure}
%\includegraphics[width=100mm]{Figures/static_eos.eps}
%\vspace{5mm}
% \caption{Mass-radius relation of nonrotating equilibrium models
%constructed with various high-density EoS, in combination with
%the two choices for the crust EoS. Individual models, differing
%by a spacing of $\Delta M=0.2 M_\odot$ in gravitational mass,
%are shown with symbols.}
%  \label{Fig:EoS}
%\end{figure}

The linearized induction equation in the MHD approximation
relates the perturbation in the magnetic field to ${\cal Y}(r)$.
Neglecting magnetic-field-induced
 $\ell \pm 2$ couplings leads to the eigenvalue equation
\begin{eqnarray}
 \Bigg[\mu + (1 + 2 \lambda_1)\frac{{a_1}^2}{\pi r^4}\Bigg]{\cal Y}''
    &+& \left\{\left(\frac{4}{r} + \Phi' - \Lambda'\right)\mu + \mu'
     + (1 + 2\lambda_1)\frac{a_1}{\pi r^4}\left[\left(\Phi'
- \Lambda'\right)a_1
+ 2{a_1}'\right]\right\}{\cal Y}'\nonumber \\
    &+& \Bigg\{\left[\left(\epsilon + p +
(1 +2\lambda_1)\frac{{a_1}^2}{\pi r^4}
\right)e^{2\Lambda}
     - \frac{\lambda_1 {{a_1}'}^2}{2\pi r^2}\right]\omega^2 e^{-2\Phi}
\nonumber \\
    &&\hspace{1cm} - (\lambda-2)\left(\frac{ \mu e^{2\Lambda}}{r^2}
- \frac{\lambda_1{{a_1}'}^2}{2\pi r^4}\right)
     + (2 + 5\lambda_1)\frac{a_1}{2\pi r^4}\left\{\left(\Phi'
- \Lambda'\right){a_1}' + {a_1}''\right\}
     \Bigg\}{\cal Y} = 0, \label{system1}
\end{eqnarray}
where $\mu$ is the shear modulus in the crust and
\begin{eqnarray}
 \lambda &=& \ell(\ell+1), \\
 \lambda_1 &=& -\frac{\ell(\ell+1)}{(2\ell-1)(2\ell+3)}.
\end{eqnarray}
In order to solve a system of first-order ODEs, we define new
variables ${\cal Y}_1$ and ${\cal Y}_2$, through
\begin{eqnarray}
 {\cal Y}_1 &\equiv& {\cal Y} r^{1-\ell}, \label{Y1} \\
 {\cal Y}_2 &\equiv& \left[\mu + (1 +2 \lambda_1)
\frac{{a_1}^2}{\pi r^4}\right]e^{\Phi-\Lambda}{{\cal Y}}'r^{2-\ell}.
    \label{Y2}
\end{eqnarray}

\begin{figure}
\includegraphics[width=80mm]{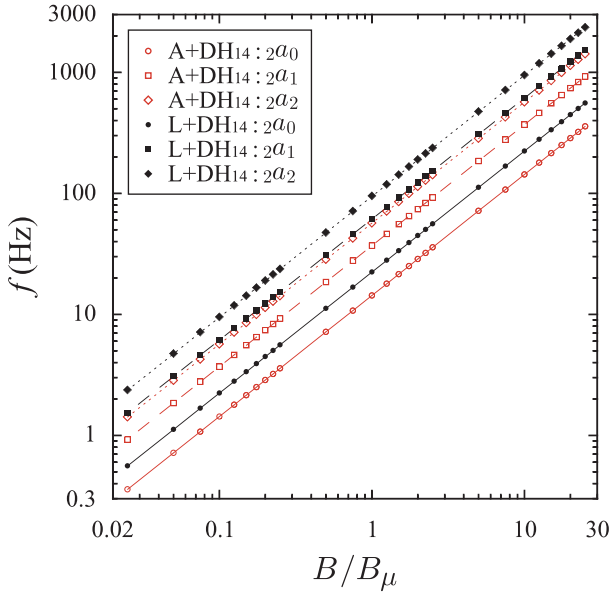}
%\vspace{5mm}
 \caption{Frequencies of Alfv\'{e}n modes, ${}_{\ell}a_n$, with
   $\ell=2$ and $n=0$, 1, and 2 as a function of the normalized
   magnetic field $B/B_\mu$.  The equilibrium models have mass
   1.4$M_\odot$ and only results for a very soft (A) and a very stiff
   (L) high-density EOS are shown, combined with the DH EOS for the
   crust (see text).  Individual numerical results are shown with
   various marks, while the continuous lines correspond to the
   empirical formula (\ref{eq:fit_l2}) with coefficient values in
   shown in Tables \ref{Tab:fit_factors0}, \ref{Tab:fit_factors1}, and
   \ref{Tab:fit_factors2}. }
  \label{Fig:FittingAL-DH14}
\end{figure}

In defining the new variables, we took into account the
form of the eigenfunction ${\cal Y}$ near the center,
where ${\cal Y}\sim r^{\ell-1}$. The final first-order system
of equations to be solved numerically for the real eigenvalues $\omega$
is then
\begin{eqnarray}
 {{\cal Y}_1}' &=& -\frac{\ell-1}{r}{\cal Y}_1
     + \frac{\pi r^3}{\pi r^4 \mu + (1 + 2\lambda_1) {a_1}^2}
e^{-\Phi + \Lambda}{\cal Y}_2,\label{eq:master3a} \\
 {{\cal Y}_2}' &=& -\Bigg[\left(\epsilon + p + (1 + 2\lambda_1)
\frac{{a_1}^2}{\pi r^4}
     - \lambda_1 e^{-2\Lambda}\frac{ {{a_1}'}^2}{2\pi r^2} \right)\omega^2
r e^{2(\Lambda-\Phi)}  \nonumber \\
    &&\hspace{1cm}- (\lambda-2)\left(\frac{ \mu e^{2\Lambda}}{r}
- \frac{\lambda_1{{a_1}'}^2}{2\pi r^3}\right)
     + (2+ 5\lambda_1) \frac{a_1 e^{2\Lambda}}{\pi r^3}
\left(\frac{a_1}{r^2} - 2\pi j_1\right)
       \Bigg]e^{\Phi - \Lambda}{\cal Y}_1
    - \frac{\ell+2}{r}{\cal Y}_2, \label{eq:master3b}
\end{eqnarray}
where the unperturbed Maxwell's equations were used in order to eliminate
the term of ${a_1}''$.

The above system of equations is solved as an eigenvalue problem,
imposing a) regularity at the center, b) a continuous traction condition
at the crust-core interface, and c) a zero-torque condition
at the stellar surface. Regularity at the center implies
\begin{eqnarray}
 {\cal Y}_2 = (\ell-1)\left[\mu + (1 +2 \lambda_1)
\frac{{\alpha_c}^2}{\pi}\right]e^{\Phi}{{\cal Y}_1}\, .
\end{eqnarray}
The demand that the traction is continuous at the crust-core
interface implies  
\begin{eqnarray}
 {\cal Y}^{'(-)} = \left[1+\frac{1}{1+2\lambda_1}
\frac{u_s^2}{u_A^2}\right]{\cal Y}^{'(+)} .
\label{trac}
\end{eqnarray}
where $u_s$ and $u_A$ are the shear and Alfv\'{e}n velocities,
respectively. 
With our choice of variables, this condition becomes
${\cal Y}_2^{(-)}={\cal Y}_2^{(+)}$. 
At the stellar surface, the zero-torque condition
$\delta T^{(s)r}_{\ \ \phi}=0$ \citep{Schumaker1983} implies
${\cal Y}_2=0$. We neglect the possible presence of a thin fluid
ocean.

Specific frequencies of predominantly or pure crustal modes are  labeled as
${}_{\ell}t_n$, while the frequencies of predominantly or pure
Alv\'{e}n modes are labeled as ${}_{\ell}a_n$, where $\ell$ is the
angular index and $n$ is associated with the number of radial nodes
in the  eigenfunction of various overtones.

\begin{table*}
 \centering
 \begin{minipage}{80mm}
  \caption{The values for the fitting factors ${}_\ell\beta_0$ (Hz)
  of equation  (\ref{eq:fit_l2}).
  The fitting factors have been calculated for magnetic field strengths up to $10^{17}$ G.}
\label{Tab:fit_factors0}
  \begin{tabular}{@{}lrrrrrrrr@{}}
  \hline
   Model  & $_2\beta_0$ &  $_3\beta_0$ &$_4\beta_0$ &  $_5\beta_0$
    & $_6\beta_0$ &  $_7\beta_0$ & $_8\beta_0$ \\
 \hline
 A+DH$_{14}$    & 14.32 & 17.67 & 20.28 & 22.66 & 24.91 & 27.08 & 29.18 \\
 A+DH$_{16}$    & 10.83 & 13.41 & 15.47 & 17.34 & 19.12 & 20.83 & 22.48 \\
 WFF3+DH$_{14}$ & 16.81 & 20.70 & 23.73 & 26.46 & 29.06 & 31.55 & 33.97 \\
 WFF3+DH$_{16}$ & 13.85 & 17.08 & 19.61 & 21.90 & 24.08 & 26.18 & 28.21 \\
 WFF3+DH$_{18}$ & 10.73 & 13.29 & 15.32 & 17.18 & 18.95 & 20.65 & 22.29 \\
 APR+DH$_{14}$  & 19.17 & 23.62 & 27.06 & 30.16 & 33.10 & 35.93 & 38.67 \\
 APR+DH$_{16}$  & 16.34 & 20.14 & 23.09 & 25.76 & 28.29 & 30.73 & 33.09 \\
 APR+DH$_{18}$  & 13.94 & 17.20 & 19.74 & 22.04 & 24.23 & 26.34 & 28.38 \\
 APR+DH$_{20}$  & 11.79 & 14.57 & 16.75 & 18.74 & 20.63 & 22.45 & 24.21 \\
 APR+DH$_{22}$  &  9.70 & 12.03 & 13.54 & 15.55 & 17.16 & 18.70 & 20.19 \\
 L+DH$_{14}$    & 22.33 & 27.44 & 31.36 & 34.87 & 38.19 & 41.38 & 44.48 \\
 L+DH$_{16}$    & 19.58 & 24.04 & 27.47 & 30.54 & 33.44 & 36.24 & 38.95 \\
 L+DH$_{18}$    & 17.29 & 21.22 & 24.24 & 26.96 & 29.53 & 32.01 & 34.42 \\
 L+DH$_{20}$    & 15.29 & 18.77 & 21.45 & 23.87 & 26.15 & 28.36 & 30.52 \\
 L+DH$_{22}$    & 13.48 & 16.57 & 18.95 & 21.10 & 23.14 & 25.12 & 27.04 \\
 L+DH$_{24}$    & 11.78 & 14.50 & 16.60 & 18.52 & 20.34 & 22.10 & 23.81 \\
 L+DH$_{26}$    &  9.97 & 12.31 & 14.14 & 15.82 & 17.42 & 18.96 & 20.46 \\
\hline
 A+NV$_{14}$    & 14.26 & 17.59 & 20.19 & 22.55 & 24.79 & 26.95 & 29.03 \\
 A+NV$_{16}$    & 10.79 & 13.37 & 15.41 & 17.27 & 19.04 & 20.75 & 22.39 \\
 WFF3+NV$_{14}$ & 16.78 & 20.67 & 23.68 & 26.41 & 29.00 & 31.49 & 33.90 \\
 WFF3+NV$_{16}$ & 13.83 & 17.06 & 19.58 & 21.87 & 24.05 & 26.14 & 28.17 \\
 WFF3+NV$_{18}$ & 10.72 & 13.28 & 15.31 & 17.16 & 18.93 & 20.63 & 22.26 \\
 APR+NV$_{14}$  & 18.47 & 22.72 & 26.00 & 28.95 & 31.74 & 34.43 & 37.04 \\
 APR+NV$_{16}$  & 15.81 & 19.46 & 22.28 & 24.82 & 27.23 & 29.56 & 31.82 \\
 APR+NV$_{18}$  & 13.55 & 16.69 & 19.12 & 21.33 & 23.43 & 25.45 & 27.41 \\
 APR+NV$_{20}$  & 11.50 & 14.19 & 16.30 & 18.21 & 20.03 & 21.78 & 23.48 \\
 APR+NV$_{22}$  &  9.49 & 11.76 & 13.54 & 15.17 & 16.72 & 18.22 & 19.67 \\
 L+NV$_{14}$    & 20.17 & 24.82 & 28.40 & 31.60 & 34.62 & 37.52 & 40.33 \\
 L+NV$_{16}$    & 17.72 & 21.80 & 24.92 & 27.71 & 30.35 & 32.89 & 35.36 \\
 L+NV$_{18}$    & 15.69 & 19.27 & 22.02 & 24.48 & 26.81 & 29.06 & 31.24 \\
 L+NV$_{20}$    & 13.91 & 17.08 & 19.51 & 21.69 & 23.76 & 25.75 & 27.70 \\
 L+NV$_{22}$    & 12.32 & 15.12 & 17.27 & 19.21 & 21.04 & 22.82 & 24.56 \\
 L+NV$_{24}$    & 10.81 & 13.27 & 15.17 & 16.88 & 18.52 & 20.10 & 21.64 \\
 L+NV$_{26}$    &  9.15 & 11.27 & 12.92 & 14.42 & 15.86 & 17.24 & 18.59 \\
\hline
\end{tabular}
\end{minipage}
\end{table*}

\begin{table*}
 \centering
 \begin{minipage}{80mm}
  \caption{The values for the fitting factors $_\ell\beta_1$ (Hz)
  of equation  (\ref{eq:fit_l2}).
  The fitting factors have been calculated for magnetic field strengths up to $10^{17}$ G.}
\label{Tab:fit_factors1}
  \begin{tabular}{@{}lrrrrrrrr@{}}
  \hline
   Model  & $_2\beta_1$ &  $_3\beta_1$ &$_4\beta_1$ &  $_5\beta_1$
    & $_6\beta_1$ &  $_7\beta_1$ & $_8\beta_1$ \\
 \hline
 A+DH$_{14}$    & 37.02 & 42.91 & 47.26 & 50.94 & 54.24 & 57.28 & 60.15 \\
 A+DH$_{16}$    & 26.71 & 31.06 & 34.30 & 37.08 & 39.59 & 41.92 & 44.14 \\
 WFF3+DH$_{14}$ & 44.50 & 51.45 & 56.54 & 60.83 & 64.65 & 68.17 & 71.48 \\
 WFF3+DH$_{16}$ & 35.80 & 41.48 & 45.67 & 49.22 & 52.39 & 55.32 & 58.08 \\
 WFF3+DH$_{18}$ & 26.46 & 30.79 & 34.04 & 36.81 & 39.32 & 41.65 & 43.87 \\
 APR+DH$_{14}$  & 51.63 & 59.53 & 65.28 & 70.12 & 74.42 & 78.39 & 82.11 \\
 APR+DH$_{16}$  & 43.51 & 50.27 & 55.21 & 59.37 & 63.09 & 66.51 & 69.73 \\
 APR+DH$_{18}$  & 36.56 & 42.32 & 46.55 & 50.13 & 53.34 & 56.30 & 59.09 \\
 APR+DH$_{20}$  & 30.30 & 35.15 & 38.74 & 41.79 & 44.53 & 47.07 & 49.47 \\
 APR+DH$_{22}$  & 24.18 & 28.14 & 30.43 & 33.63 & 35.92 & 38.05 & 40.08 \\
 L+DH$_{14}$    & 61.38 & 70.83 & 77.71 & 83.48 & 88.60 & 93.29 & 97.66 \\
 L+DH$_{16}$    & 53.30 & 61.56 & 67.57 & 72.61 & 77.08 & 81.17 & 84.99 \\
 L+DH$_{18}$    & 46.51 & 53.78 & 59.08 & 63.53 & 67.47 & 71.08 & 74.45 \\
 L+DH$_{20}$    & 40.55 & 46.93 & 51.61 & 55.54 & 59.03 & 62.23 & 65.22 \\
 L+DH$_{22}$    & 35.18 & 40.74 & 44.84 & 48.29 & 51.36 & 54.19 & 56.84 \\
 L+DH$_{24}$    & 30.13 & 34.93 & 38.47 & 41.46 & 44.14 & 46.61 & 48.94 \\
 L+DH$_{26}$    & 24.75 & 28.77 & 31.75 & 34.29 & 36.57 & 38.69 & 40.70 \\
\hline
 A+NV$_{14}$    & 36.88 & 42.75 & 47.07 & 50.74 & 54.02 & 57.05 & 59.90 \\
 A+NV$_{16}$    & 26.62 & 30.95 & 34.19 & 36.95 & 39.45 & 41.77 & 43.98 \\
 WFF3+NV$_{14}$ & 44.43 & 51.37 & 56.45 & 60.73 & 64.55 & 68.06 & 71.36 \\
 WFF3+NV$_{16}$ & 35.77 & 41.44 & 45.63 & 49.16 & 52.33 & 55.26 & 58.02 \\
 WFF3+NV$_{18}$ & 26.43 & 30.76 & 34.00 & 36.77 & 39.27 & 41.60 & 43.82 \\
 APR+NV$_{14}$  & 49.79 & 57.43 & 63.00 & 67.66 & 71.81 & 75.61 & 79.18 \\
 APR+NV$_{16}$  & 42.12 & 48.66 & 53.44 & 57.46 & 61.04 & 64.33 & 67.42 \\
 APR+NV$_{18}$  & 35.53 & 41.13 & 45.23 & 48.70 & 51.79 & 54.64 & 57.32 \\
 APR+NV$_{20}$  & 29.56 & 34.29 & 37.78 & 40.74 & 43.39 & 45.85 & 48.16 \\
 APR+NV$_{22}$  & 23.68 & 27.54 & 30.43 & 32.90 & 35.13 & 37.20 & 39.16 \\
 L+NV$_{14}$    & 56.55 & 65.42 & 71.91 & 77.35 & 82.18 & 86.59 & 90.69 \\
 L+NV$_{16}$    & 49.07 & 56.88 & 62.60 & 67.40 & 71.65 & 75.52 & 79.12 \\
 L+NV$_{18}$    & 42.73 & 49.61 & 54.67 & 58.92 & 62.69 & 66.13 & 69.32 \\
 L+NV$_{20}$    & 37.20 & 43.21 & 47.65 & 51.40 & 54.73 & 57.77 & 60.60 \\
 L+NV$_{22}$    & 32.28 & 37.50 & 41.36 & 44.62 & 47.52 & 50.18 & 52.67 \\
 L+NV$_{24}$    & 27.68 & 32.15 & 35.45 & 38.25 & 40.75 & 43.05 & 45.20 \\
 L+NV$_{26}$    & 22.68 & 26.39 & 29.15 & 31.49 & 33.58 & 35.52 & 37.34 \\
\hline
\end{tabular}
\end{minipage}
\end{table*}

\begin{table*}
 \centering
 \begin{minipage}{80mm}
  \caption{The values for the fitting factors $_\ell\beta_2$ (Hz)
  of equation  (\ref{eq:fit_l2}).
  The fitting factors have been calculated for magnetic field strengths up to $10^{17}$ G.}
\label{Tab:fit_factors2}
  \begin{tabular}{@{}lrrrrrrrr@{}}
  \hline
   Model  & $_2\beta_2$ &  $_3\beta_2$ &$_4\beta_2$ &  $_5\beta_2$
    & $_6\beta_2$ &  $_7\beta_2$ & $_8\beta_2$ \\
 \hline
 A+DH$_{14}$    & 56.74 & 64.60 & 70.36 & 75.22 & 79.55 & 83.50 & 87.16 \\
 A+DH$_{16}$    & 40.47 & 46.17 & 50.39 & 53.96 & 57.16 & 60.10 & 62.83 \\
 WFF3+DH$_{14}$ & 68.12 & 77.39 & 84.13 & 89.80 & 94.82 & 99.40 &103.64 \\
 WFF3+DH$_{16}$ & 54.59 & 62.10 & 67.60 & 72.23 & 76.34 & 80.10 & 83.59 \\
 WFF3+DH$_{18}$ & 40.06 & 45.70 & 49.88 & 53.44 & 56.61 & 59.53 & 62.25 \\
 APR+DH$_{14}$  & 78.92 & 89.48 & 97.12 &103.52 &109.19 &114.35 &119.14 \\
 APR+DH$_{16}$  & 66.42 & 75.38 & 81.87 & 87.32 & 92.14 & 96.54 &100.62 \\
 APR+DH$_{18}$  & 55.73 & 63.33 & 68.87 & 73.53 & 77.67 & 81.44 & 84.95 \\
 APR+DH$_{20}$  & 46.07 & 52.45 & 57.13 & 61.08 & 64.60 & 67.81 & 70.81 \\
 APR+DH$_{22}$  & 36.59 & 41.75 & 44.62 & 48.82 & 51.73 & 54.40 & 56.89 \\
 L+DH$_{14}$    & 94.61 &107.36 &116.59 &124.31 &131.13 &137.33 &143.05 \\
 L+DH$_{16}$    & 81.89 & 93.02 &101.10 &107.86 &113.83 &119.25 &124.25 \\
 L+DH$_{18}$    & 71.30 & 81.00 & 88.07 & 94.02 & 99.28 &104.06 &108.48 \\
 L+DH$_{20}$    & 62.13 & 70.60 & 76.76 & 81.95 & 86.55 & 90.74 & 94.62 \\
 L+DH$_{22}$    & 53.86 & 61.24 & 66.62 & 71.15 & 75.15 & 78.81 & 82.19 \\
 L+DH$_{24}$    & 46.01 & 52.38 & 57.03 & 60.96 & 64.44 & 67.61 & 70.56 \\
 L+DH$_{26}$    & 37.48 & 42.79 & 46.71 & 50.03 & 52.98 & 55.68 & 58.18 \\
\hline
 A+NV$_{14}$    & 56.53 & 64.35 & 70.09 & 74.93 & 79.24 & 83.17 & 86.82 \\
 A+NV$_{16}$    & 40.34 & 46.02 & 50.22 & 53.78 & 56.97 & 59.89 & 62.61 \\
 WFF3+NV$_{14}$ & 68.02 & 77.27 & 84.00 & 89.66 & 94.67 & 99.24 &103.47 \\
 WFF3+NV$_{16}$ & 54.54 & 62.04 & 67.53 & 72.16 & 76.26 & 80.02 & 83.50 \\
 WFF3+NV$_{18}$ & 40.01 & 45.65 & 49.83 & 53.38 & 56.55 & 59.46 & 62.19 \\
 APR+NV$_{14}$  & 76.27 & 86.53 & 93.97 &100.20 &105.72 &110.74 &115.39 \\
 APR+NV$_{16}$  & 64.38 & 73.09 & 79.41 & 84.71 & 89.40 & 93.66 & 97.62 \\
 APR+NV$_{18}$  & 54.21 & 61.61 & 67.01 & 71.54 & 75.56 & 79.22 & 82.62 \\
 APR+NV$_{20}$  & 44.97 & 51.19 & 55.76 & 59.61 & 63.03 & 66.16 & 69.07 \\
 APR+NV$_{22}$  & 35.84 & 40.88 & 44.62 & 47.80 & 50.63 & 53.23 & 55.66 \\
 L+NV$_{14}$    & 86.88 & 98.86 &107.60 &114.91 &121.37 &127.24 &132.65 \\
 L+NV$_{16}$    & 75.15 & 85.56 & 93.22 & 99.69 &105.43 &110.64 &115.46 \\
 L+NV$_{18}$    & 65.52 & 74.55 & 81.18 & 86.81 & 91.82 & 96.40 &100.65 \\
 L+NV$_{20}$    & 57.20 & 65.09 & 70.87 & 75.75 & 80.08 & 84.05 & 87.73 \\
 L+NV$_{22}$    & 49.67 & 56.58 & 61.64 & 65.91 & 69.69 & 73.13 & 76.33 \\
 L+NV$_{24}$    & 42.46 & 48.42 & 52.80 & 56.50 & 59.79 & 62.78 & 65.55 \\
 L+NV$_{26}$    & 34.41 & 39.34 & 43.01 & 46.11 & 48.87 & 51.39 & 53.73 \\
\hline
\end{tabular}
\end{minipage}
\end{table*}

%%%%%%%%%%%%%%%%%%%%%%%%%%%%%%%%%%%%%%%%%%%%%%%%%%%%%%%%%%%%%%%%%%%%%%%%%%%%
%%%%%%%%%%%%%%%%%%%%%%%%%%%%%%%%%%%%%%%%%%%%%%%%%%%%%%%%%%%%%%%%%%%%%%%%%%%%
\section{Numerical Results}
\label{sec:III}
%%%%%%%%%%%%%%%%%%%%%%%%%%%%%%%%%%%%%%%%%%%%%%%%%%%%%%%%%%%%%%%%%%%%%%%%%%%%
%%%%%%%%%%%%%%%%%%%%%%%%%%%%%%%%%%%%%%%%%%%%%%%%%%%%%%%%%%%%%%%%%%%%%%%%%%%%

Our equilibrium models comprise both a fluid part and a thin
solid crust, which can have a thickness ranging from about 3\%
to 12\% of the stellar radius, depending on the EOS and mass.
We will compute global torsional Alfv\'{e}n modes in the
presence of both of these components. However, in order to
understand qualitatively the properties of these modes, we
will first study them in the limiting case where one neglects
the presence of a solid crust and treats the model as a pure
magneto-fluid. In addition, in order to understand the influence
of the presence of a solid crust on the Alfv\'{e}n modes (and,
conversely, the influence of a global magneto-fluid on the
torsional crust modes) we will study a toy model of varying
crust thickness, before focusing on the realistic models
presented in the previous section.

%%%%%%%%%%%%%%%%%%%%%%%%%%%%%%%%%%%%%%%%%%%%%%%%%%%%%%%%%%%%%%%%%%%%%%%%%%%%
\subsection{Global Alfv\'{e}n modes for a pure magneto-fluid}
%%%%%%%%%%%%%%%%%%%%%%%%%%%%%%%%%%%%%%%%%%%%%%%%%%%%%%%%%%%%%%%%%%%%%%%%%%%%

As a limiting case, we first consider global Alfv\'{e}n modes, in the
absence of a solid crust, i.e. setting $\mu=0$ in the crust region.
These modes are pure Alfv\'{e}n modes and degenerate to
zero frequency in the limit of a vanishing magnetic field, since then
the eigenvalue problem becomes
\begin{equation}
 (\epsilon + p)e^{-2\Phi + 2\Lambda}\omega^2 {\cal Y}=0,
\end{equation}
which has $\omega=0$ as a trivial solution. In the presence of
a magnetic field, the eigenfrequency becomes proportional to the
magnetic field strength $B$.

Figure \ref{Fig:FittingAL-DH14} shows the frequency of several modes
(with $\ell=2$ and $n=0,1,2$) as a function of the strength of the
magnetic field, for $1.4 M_\odot$ models of a very soft (A) and a very
stiff (L) EOS.  Various lines in the figure are linear fits to the
numerical data. As is apparent from these numerical results, the
linearity of the frequency as a function of $B$ is preserved to a very
high accuracy, at least in the range of $B=10^{14}$ -- $10^{17}$ G.
Therefore, our numerical results can be represented by the following
 empirical formula
\begin{equation}
 {}_{\ell}a_{n} = {}_{\ell}\beta_n \left(\frac{B}{B_{\mu}}\right),
\label{eq:fit_l2}
\end{equation}
where $B_{\mu}$ is a typical magnetic field strength, which we take to
be $B_{\mu}=4\times 10^{15}$ G (as in Paper I) and ${}_{\ell}\beta_n$
are coefficients, which depend on the equilibrium model and the mode
in question.  For each equilibrium stellar model the coefficients
${}_{\ell}\beta_n$ for $\ell=2-8$ and $n=0,1,2$ are shown in Tables
\ref{Tab:fit_factors0}, \ref{Tab:fit_factors1}, and
\ref{Tab:fit_factors2}.

\begin{figure}
\includegraphics[width=80mm]{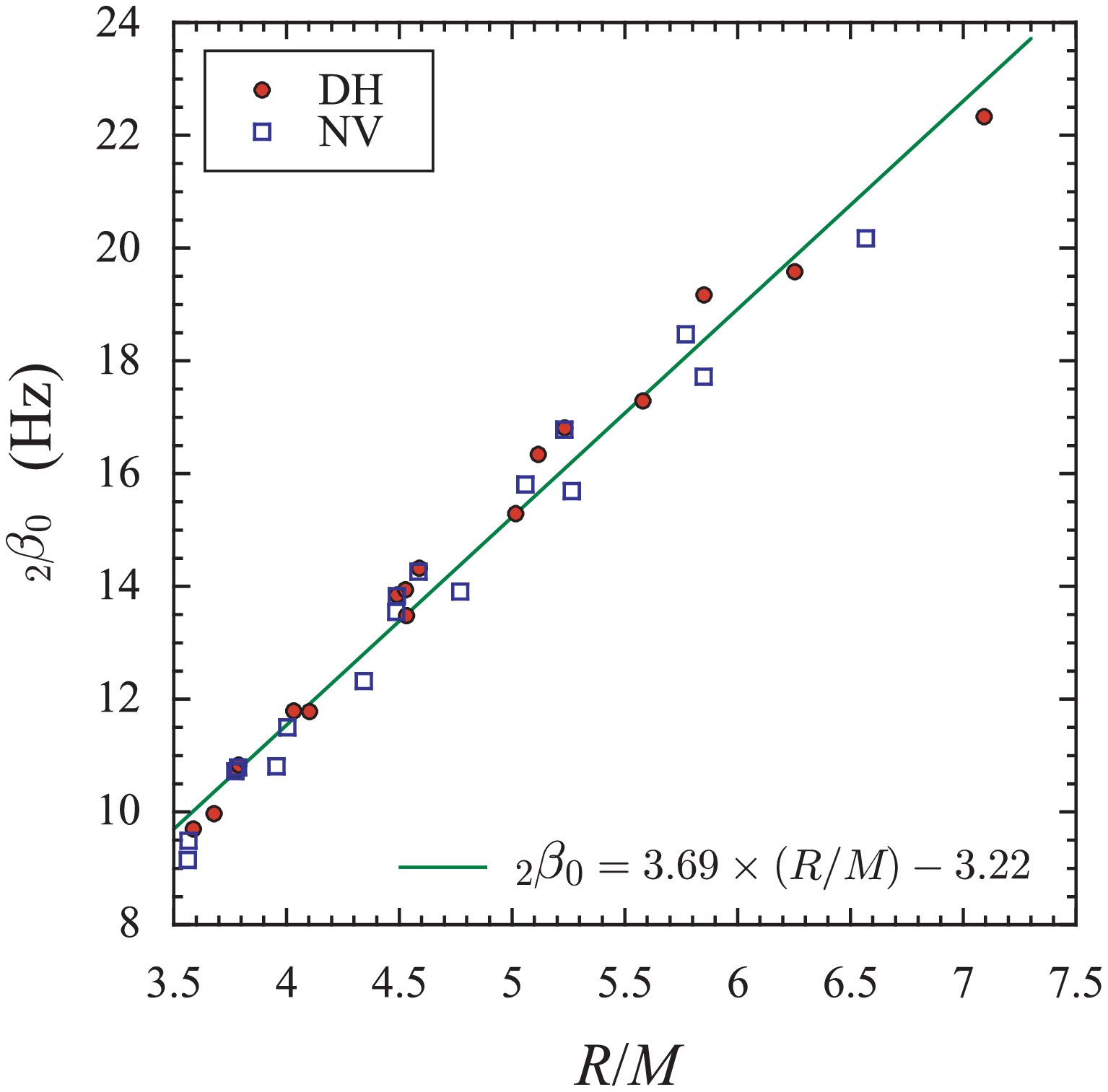}
%\vspace{5mm}
 \caption{Distribution of the coefficients ${}_{2}\beta_0$
   in Table \ref{Tab:fit_factors0}
   with respect to the inverse of stellar compactness $M/R$. The marks
   of circle and square correspond to the crust EOS DH and NV,
   respectively, while the solid line is a linear fit, used in
   constructing Equation (\ref{roverm}). }
  \label{Fig:a-compactness}
\end{figure}

As a general trend, one sees that the value of the coefficients
${}_{\ell}\beta_n$ becomes larger (and thus the dependence of the
mode frequency on the magnetic field strength becomes stronger), as
the mass of the star increases (for a given EOS). The same trend holds
with increasing stiffness of the EOS, for a given mass.  One also sees
from the Table \ref{Tab:fit_factors0} that
${}_{\ell}\beta_0/{}_{2}\beta_0\approx\sqrt{\ell/2}$, so that the
following empirical formula holds for the fundamental modes
\begin{equation}
 {}_{\ell}a_0 \approx {}_{2}\beta_0 \sqrt{\frac{\ell}{2}}
\left(\frac{B}{B_{\mu}}\right).  \label{eq:fit_l3}
\end{equation}
Examining more closely the dependence of the mode frequency on the
parameters of the equilibrium models, we find that the coefficients
${}_{\ell}\beta_0$ are a linear function of the inverse of the
compactness $M/R$ (see Figure \ref{Fig:a-compactness}), so that the
frequency of the fundamental modes is approximately given by
\begin{equation}
 {}_{\ell}a_0 \approx \sqrt{\frac{\ell}{2}}
\left[3.69\left(\frac{R}{M}\right) - 3.22\right]\left(\frac{B}{B_{\mu}}\right),
\label{roverm}
\end{equation}
(with an accuracy of a few per cent).
For given strength of the magnetic field the frequency of the
fundamental modes varies by a factor of up to 2.4 within the set
equilibrium models in Table \ref{Tab:fit_factors0}. Similarly,
the frequency of the $n=1$ and $n=2$ overtones of the $\ell=2$ mode
varies by up to a factor of 2.7. It follows that the frequency
of global Alfv\'{e}n modes is sensitive to the magnetic field
strength and the bulk properties of the equilibrium model and
a successful identification of several modes would, in principle,
allow the inference of both the magnetic field and the compactness
of the star, through empirical relations such as Equation (\ref{roverm}).

\begin{figure}
\includegraphics[width=80mm]{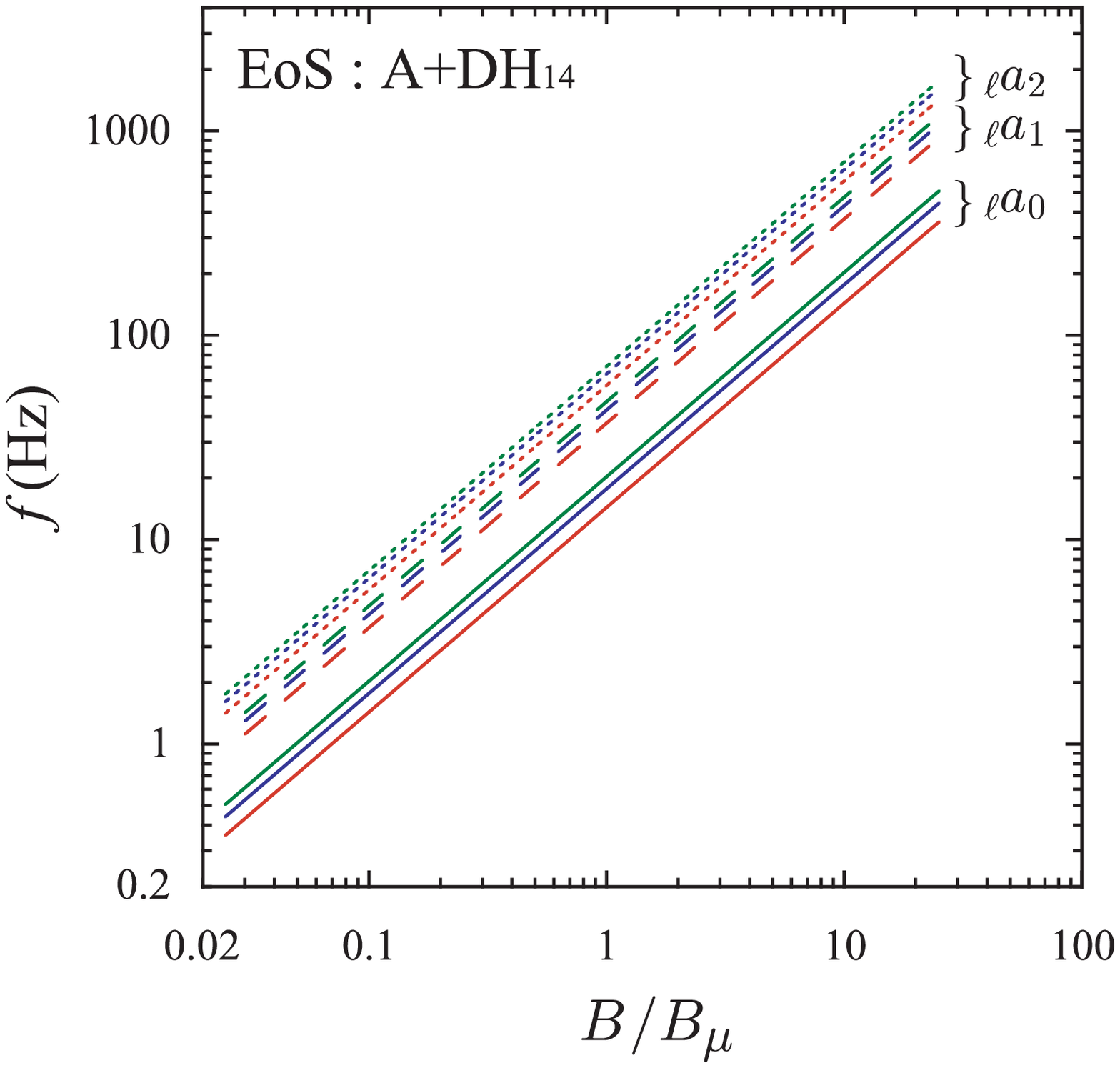}
%\vspace{5mm}
 \caption{The frequencies of Alfv\'{e}n modes for ${}_{\ell}a_n$ with
   $\ell=2$, 3, and 4 and $n=0$, 1, and 2 as functions of the
   normalized magnetic field $B/B_\mu$.  The solid,
   dashed, and short-dashed lines correspond to the frequencies of
   ${}_{\ell}a_0$, ${}_{\ell}a_1$, and ${}_{\ell}a_2$, respectively,
   and for each $n$ the frequencies increase with increasing $\ell$.
   For a given $\ell$, the frequency of various overtones does not
   coincide, in sharp contrast with crustal torsional modes.
   The mass of the equilibrium model, constructed
   with the A+DH EOS, is $M=1.4M_\odot$.  }
  \label{Fig:A_DH_n01_l234}
\end{figure}

Figure \ref{Fig:A_DH_n01_l234} shows the frequencies
with $\ell=2$, 3, and 4, and $n=0$, 1, and 2 as a function of the
magnetic field strength $B/B_{\mu}$, for the particular stellar model
A+DH$_{14}$.  The Figure demonstrates that the frequencies of overtones
depend on the value of $\ell$, in sharp contrast to the case of pure crust
modes, where the frequencies of overtones are almost
independent of $\ell$ (see, e.g. Paper I).

%%%%%%%%%%%%%%%%%%%%%%%%%%%%%%%%%%%%%%%%%%%%%%%%%%%%%%%%%%%%%%%%%%%%%%%%%%%%
\subsection{Avoided crossings between global Alfv\'{e}n
modes and torsional modes of the crust}
%%%%%%%%%%%%%%%%%%%%%%%%%%%%%%%%%%%%%%%%%%%%%%%%%%%%%%%%%%%%%%%%%%%%%%%%%%%%

How does the presence of a solid crust affect the global Alfv\'{e}n
modes?  And, conversely, how are the torsional modes of the crust
affected when global Alfv\'{e}n modes are also present? In order to
answer these questions, we construct a toy model in which we introduce
a crust of varying thickness $\Delta r / R$. We focus on an
equilibrium models with $M=1.4M_{\odot}$ and $R=10.4$ km, constructed
with a relativistic polytropic EOS with polytropic index $N=1.0$ and
polytropic constant $K=130$ km$^2$. In this toy model, the shear
modulus is simply taken to be $\mu =\rho v_s^2$, where $v_s=1.0\times
10^{8}$ cm/s \citep{Schumaker1983}.

If one assumes that the whole star is solid ($\Delta r/ R=1.0$) then
the frequency spectrum of torsional modes only consist of crust modes.
Qualitatively, the dependence of the mode frequencies on the strength
of the magnetic field is similar to the results in Paper I, i.e. there
is a weak dependence up to $B/B_\mu \sim 1$ (nearly horizontal
branch), followed by a rapid increase in frequency for larger magnetic
field strengths (magnetic-field-dominated branch).  When the crust
thickness is reduced to $\Delta r/R<1$, global Alfv\'{e}n modes are
introduced. In such a composite fluid/crust case, the eigenfunctions
of torsional crust modes penetrate inside the fluid region, i.e. crust
modes become global modes and, conversely, the eigenfunctions of
Alfv\'{e}n modes penetrate the inside crust region.  Thus, the allowed
discrete global modes are only those for which the eigenfunction at
the crust/core interface satisfies the continuous traction condition
(\ref{trac}) exactly. As we will see, this strict requirement leads to
avoided crossings between the two families of modes and to a dramatic
change of the mode spectrum in the limit of small crust thickness and
high magnetic field.

\begin{figure}
\begin{center}
\begin{tabular}{cc}
\includegraphics[scale=0.45]{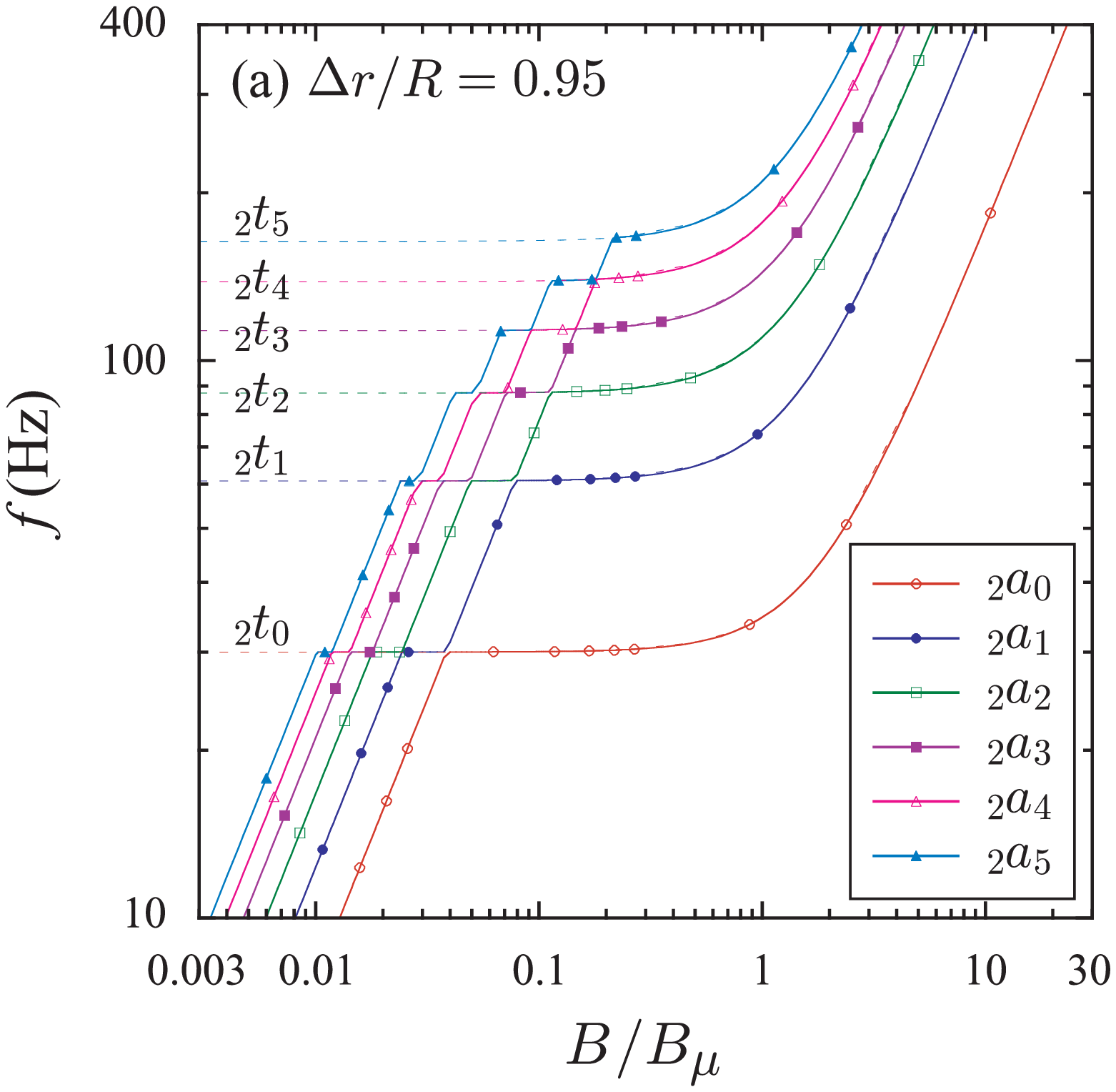} &
\includegraphics[scale=0.45]{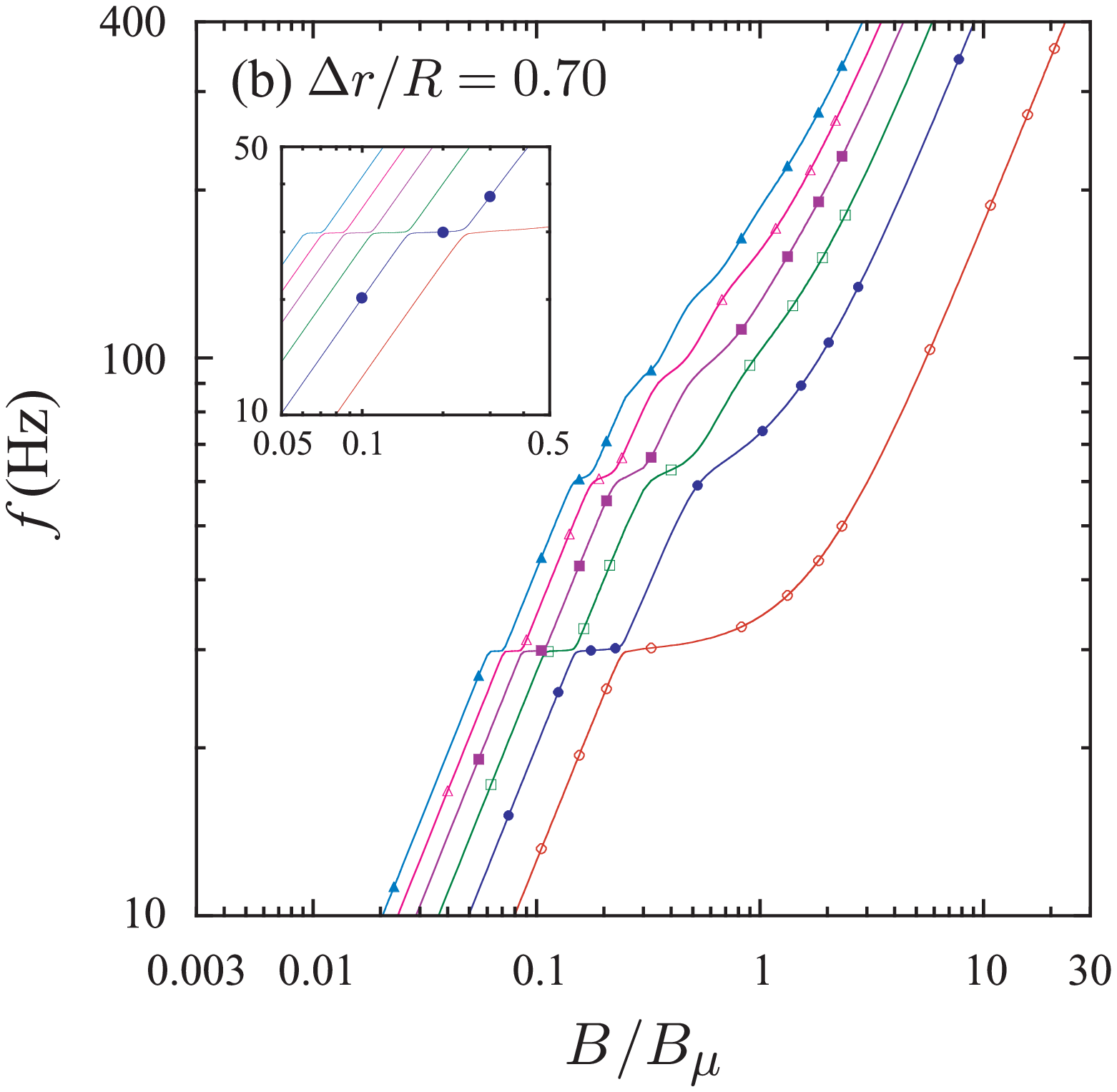} \\
\includegraphics[scale=0.45]{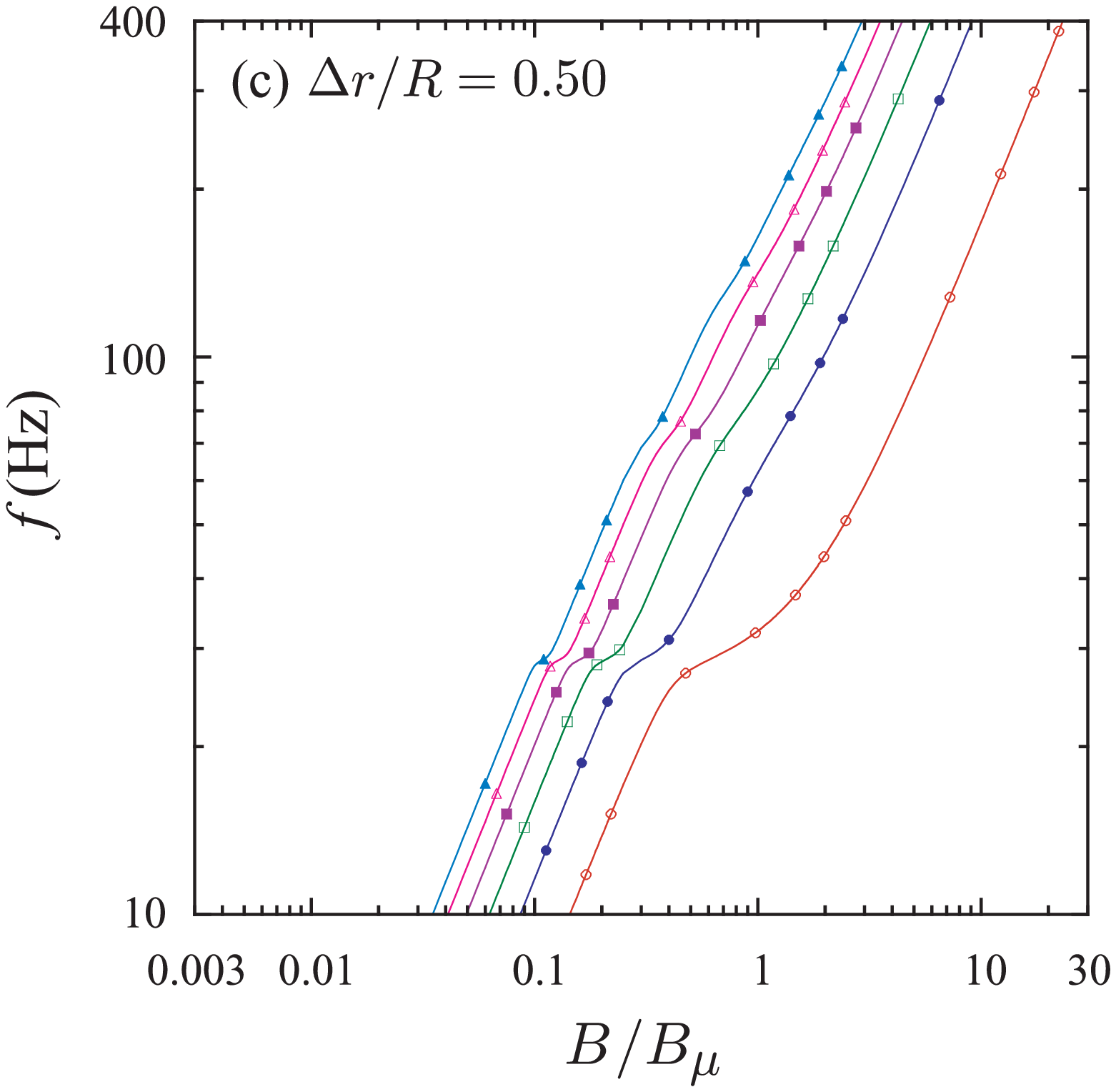} &
\includegraphics[scale=0.45]{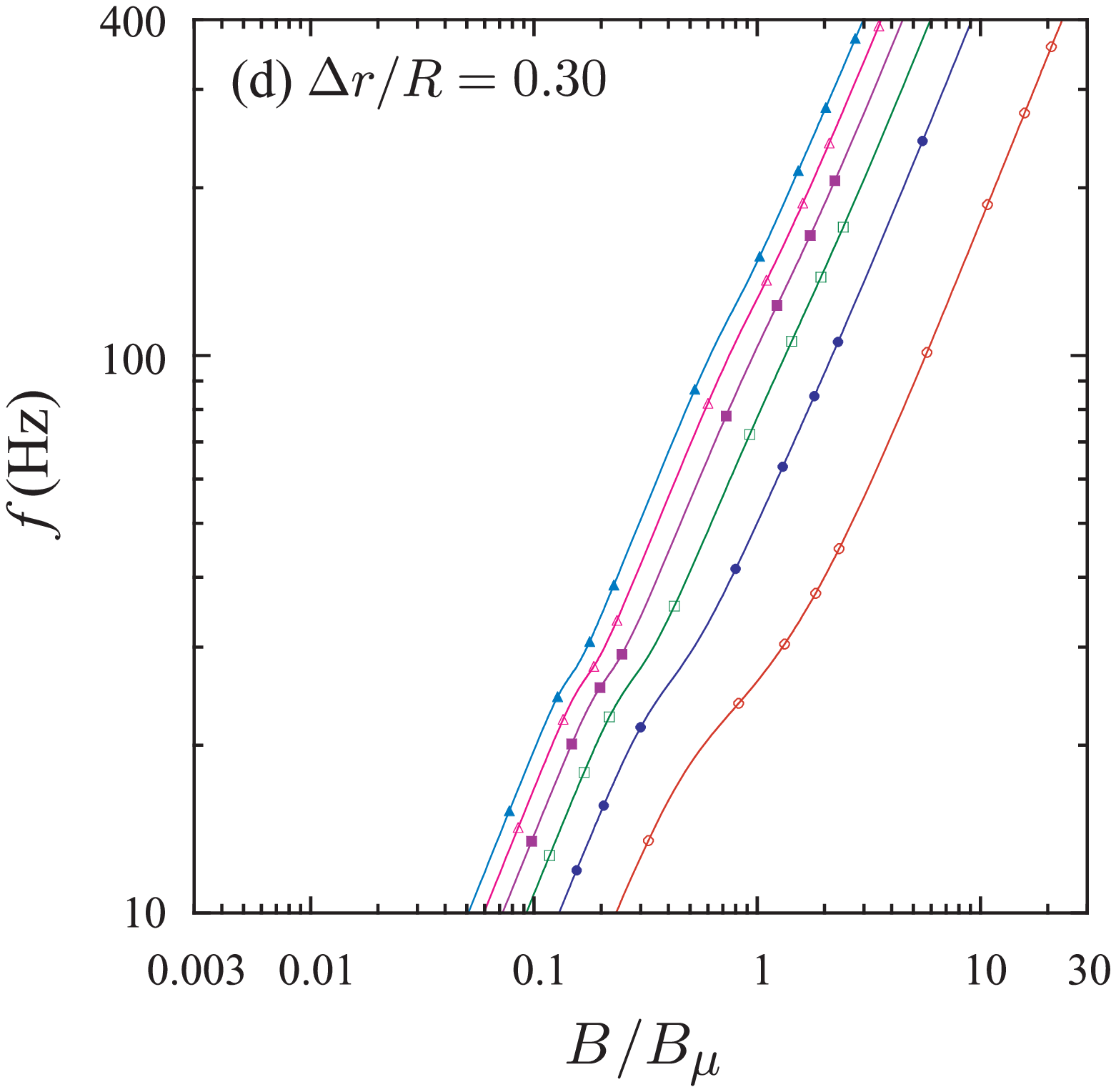}
\end{tabular}
\end{center}
\caption{The behavior of torsional crust modes $({}_{l}t_n)$
  and global Alfv\'{e}n modes $({}_{l}a_n)$ for different toy models,
  when varying the crust thickness $\Delta r/R$. The first six
  eigenfrequencies for $\ell=2$ are shown. For comparison, we also plot
  the case of $\Delta r / R=1.0$ (no fluid region) as dashed lines in
  panel (a).  }
\label{fig:behavior-thickness}
\end{figure}

Figure \ref{fig:behavior-thickness} shows the first six frequencies
for $\ell=2$, as a function of the strength of magnetic field, for
four different values of the crust thickness, $\Delta r/R=0.95$, 0.70.
0.50, and 0.30.  As seen Figure \ref{fig:behavior-thickness}(a) even a
small fluid part ($\Delta r/ R=0.95$) allows both families of modes to
be present, e.g. for small magnetic field strengths one can see both
horizontal mode sequences (dashed lines, corresponding to crust modes)
and Alfv\'{e}n-like sequences (continuous lines). However, as the
functional dependence of the global Alfv\'{e}n modes with increasing
magnetic field strength is different than the functional dependence
of crust modes, the two families of modes appear to cross each other.
On closer examination, one sees that, in fact, avoided crossings
take place. For ($\Delta r/ R=0.95$) the gap at avoided crossings
is still small, so that both families of modes are clearly visible.
In addition, since the fluid region is still small, the Alfv\'{e}n mode
frequencies are high and the avoided crossings take place on the
horizontal branch of the crust modes.

When the crust thickness is reduced to $\Delta r/R = 0.70$, Figure
\ref{fig:behavior-thickness}(b), the enlargement of the fluid region
leads to a reduction of the Alfv\'{e}n mode frequencies. This leads to
the avoided crossings taking place at larger values of the magnetic
field. The avoided crossing of the fundamental mode takes place close
to the transition of the crust modes from the horizontal branch to the
magnetic-field-dominated branch. Thus, a horizontal branch of
predominantly crust modes is still present. However, the avoided
crossings of higher-order overtones take place at the
magnetic-field-dominated branch of crust modes. This leads to a
significant increase in the gap at avoided crossings and one
cannot easily distinguish modes that are predominantly crust modes.

Reducing the crust thickness further to $\Delta r/R=0.50$ and 0.30,
Figures \ref{fig:behavior-thickness}(c),(d), one observes that the
avoided crossings of low-order modes take place in the
magnetic-field-dominated branch of crust modes and the gap in
avoided crossings tends to become of the same order as the
frequency spacing of consecutive Alfv\'{e}n modes. The avoided
crossings are now visible as only a small local change in the
slope of mode-sequences, where one would expect a crust mode to
be present. However, modes of predominantly crustal character
are no longer distinguishable. The same trend continues when the
crust thickness is reduced even further, so that for realistic
values of the crust thickness (3\%-12\%) the mode spectrum is
similar to the pure Alfv\'{e}n mode spectrum shown
in Figure \ref{Fig:A_DH_n01_l234}.

\begin{figure}
\includegraphics[width=80mm]{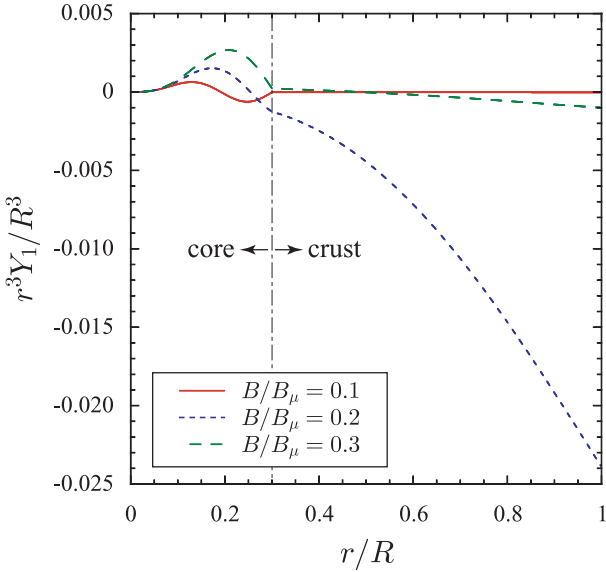}
%\vspace{5mm}
 \caption{Scaled eigenfunctions of the ${}_2a_1$ (at $B/B_{\mu}=0.1$),
  ${}_2t_0$ (at $B/B_{\mu}=0.2$) and ${}_2a_0$ (at $B/B_{\mu}=0.1$) modes
  for a toy model with large crust thickness of $\Delta r/R =0.70$. The
  number of nodes and the relative amplitude at the crust/fluid regions
  is consistent with the expected behavior for predominantly crustal
  modes and predominantly Alfv\'{e}n-like modes, respectively.}
  \label{Fig:dR70-f1}
\end{figure}

Regarding the eigenfunctions of oscillation modes, at a given magnetic
field strength, the lowest-frequency mode corresponds to a fundamental
mode without any node, while the number of nodes is increasing for
overtones of higher frequency.  We find that the position of nodes
change at avoided crossings. The inset in Figure
\ref{fig:behavior-thickness}(b) follows the continuous sequence of
mode frequencies, which start out as the ${}_2a_1$ Alfv\'{e}n mode
(first filled circle at $B/B_{\mu}=0.1$). After an avoided crossing,
the sequence becomes a fundamental torsional mode, ${}_2t_0$, (second
filled circle at $B/B_{\mu}=0.2$). After a second avoided crossing the
continuous sequence changes into fundamental Alfv\'{e}n mode,
${}_2a_0$, (third filled circle at $B/B_{\mu}=0.3$). Figure
\ref{Fig:dR70-f1} shows that each of the above three eigenfunctions
has one node. Specifically, the eigenfunctions for the ${}_2a_1$ and
 ${}_2t_0$ modes have one node in the fluid region while the ${}_2a_0$
mode has one node in the crust.

We further observe that the amplitude of the ${}_2a_0$ and ${}_2a_1$
Alfv\'{e}n modes is larger in the fluid region, while the amplitude
of the ${}_2t_0$ mode is larger in the crust region. This is
consistent with the interpretation that, even with the presence of
a large number of avoided crossings, horizontal branches correspond to
predominantly crust-like modes, while branches of increasing
frequency (with increasing magnetic field strength) correspond to
predominantly Alfv\'{e}n-like modes (except for the magnetically-dominated
branch of crust modes). Of course, in the above example a large crust
thickness of $\Delta r/R = 0.70$ was used, where the two family
of modes are still distinguishable. For a much smaller crust thickness,
the the horizontal branch of crust modes cannot be clearly identified
(due to the strong avoided crossings), while the magnetically-dominated
branch of crust modes is merged with the Alfv\'{e}n modes, as both
families of modes have nearly the same frequencies and slopes at
large magnetic field strengths.

%%%%%%%%%%%%%%%%%%%%%%%%%%%%%%%%%%%%%%%%%%%%%%%%%%%%%%%%%%%%%%%%%%%%%%%%%%%%
\subsection{Global Alfv\'{e}n modes for realistic models}
%%%%%%%%%%%%%%%%%%%%%%%%%%%%%%%%%%%%%%%%%%%%%%%%%%%%%%%%%%%%%%%%%%%%%%%%%%%%

\begin{figure}
\includegraphics[width=80mm]{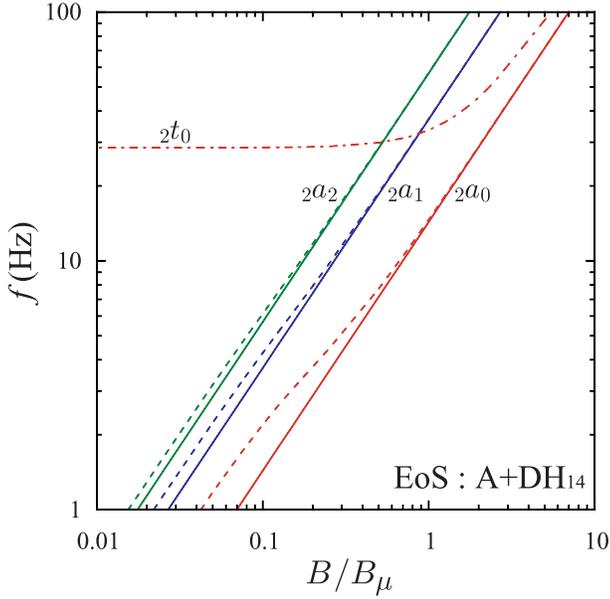}
%\vspace{5mm}
 \caption{The frequencies of global Alfv\'{e}n  modes ${}_{\ell}a_n$ with
   $\ell=2$ and $n=0$, 1, and 2 as functions of the normalized
   magnetic field ($B/B_\mu$).  The dashed line corresponds to a
   realistic model (A+DH$_{14}$) with both fluid and solid regions.
   The solid line corresponds the same model, but setting $\mu=0$ in
   the crust region (pure magneto-fluid). The differences between the
   two cases are limited to small magnetic field strengths, while for
   higher values there is coincidence. For comparison, we also show
   the ${}_{2}t_0$ pure crustal mode (when the magnetic field is
   confined to the crust).}
  \label{Fig:AGC-ADH14}
\end{figure}

Realistic stellar models have very thin solid crusts and the frequency
spectrum of global modes approaches that of the pure Alfv\'{e}n modes.
As an example, Figure \ref{Fig:AGC-ADH14} shows the three lowest-order
$\ell=2$ modes, ${}_2a_0$, ${}_2a_1$, and ${}_2a_2$ for model
A+DH$_{14}$. Pure Alfv\'{e}n modes (no solid crust) are shown as solid
lines while the modes in the presence of the crust are shown as dashed
lines. The pure crustal fundamental mode (when the magnetic field is
confined to the crust region only, see Paper I) is also shown as a
dash-dotted line. For weak magnetic fields the presence of the solid
crust has some effect on the Alfv\'{e}n mode frequencies, but, for
large magnetic fields the effect diminishes.  For larger $\ell$, the
effect of the presence of a solid crust on the mode frequencies
persists for somewhat stronger magnetic fields.

Since the global modes for large magnetic field strengths are almost
identical to pure Alfv\'{e}n modes, Equation (\ref{eq:fit_l2})
describes their frequencies with good accuracy, for large magnetic
field strengths.  In addition, the non-degeneracy of the
frequency of overtones for different values of the harmonic index
$\ell$ will be important in trying to match numerical results to
observed frequencies.

%%%%%%%%%%%%%%%%%%%%%%%%%%%%%%%%%%%%%%%%%%%%%%%%%%%%%%%%%%%%%%%%%%%%%%%%%%%%
%%%%%%%%%%%%%%%%%%%%%%%%%%%%%%%%%%%%%%%%%%%%%%%%%%%%%%%%%%%%%%%%%%%%%%%%%%%%
\section{Comparison with observed frequencies in SGRs}
\label{sec:IV}
%%%%%%%%%%%%%%%%%%%%%%%%%%%%%%%%%%%%%%%%%%%%%%%%%%%%%%%%%%%%%%%%%%%%%%%%%%%%
%%%%%%%%%%%%%%%%%%%%%%%%%%%%%%%%%%%%%%%%%%%%%%%%%%%%%%%%%%%%%%%%%%%%%%%%%%%%

As mentioned in Paper I, it is difficult to explain all of the observed
frequencies in SGRs using pure crustal torsional modes (with magnetic
fields confined to the crust only). An example is the small separation
of the two frequencies of 26 and 29 Hz for SGR 1806-20. Here, we find
that such frequency data can be easily explained with global
Alfv\'{e}n modes, owing to the non-degeneracy of overtones for
different values of $\ell$.

\begin{figure}
\includegraphics[width=80mm]{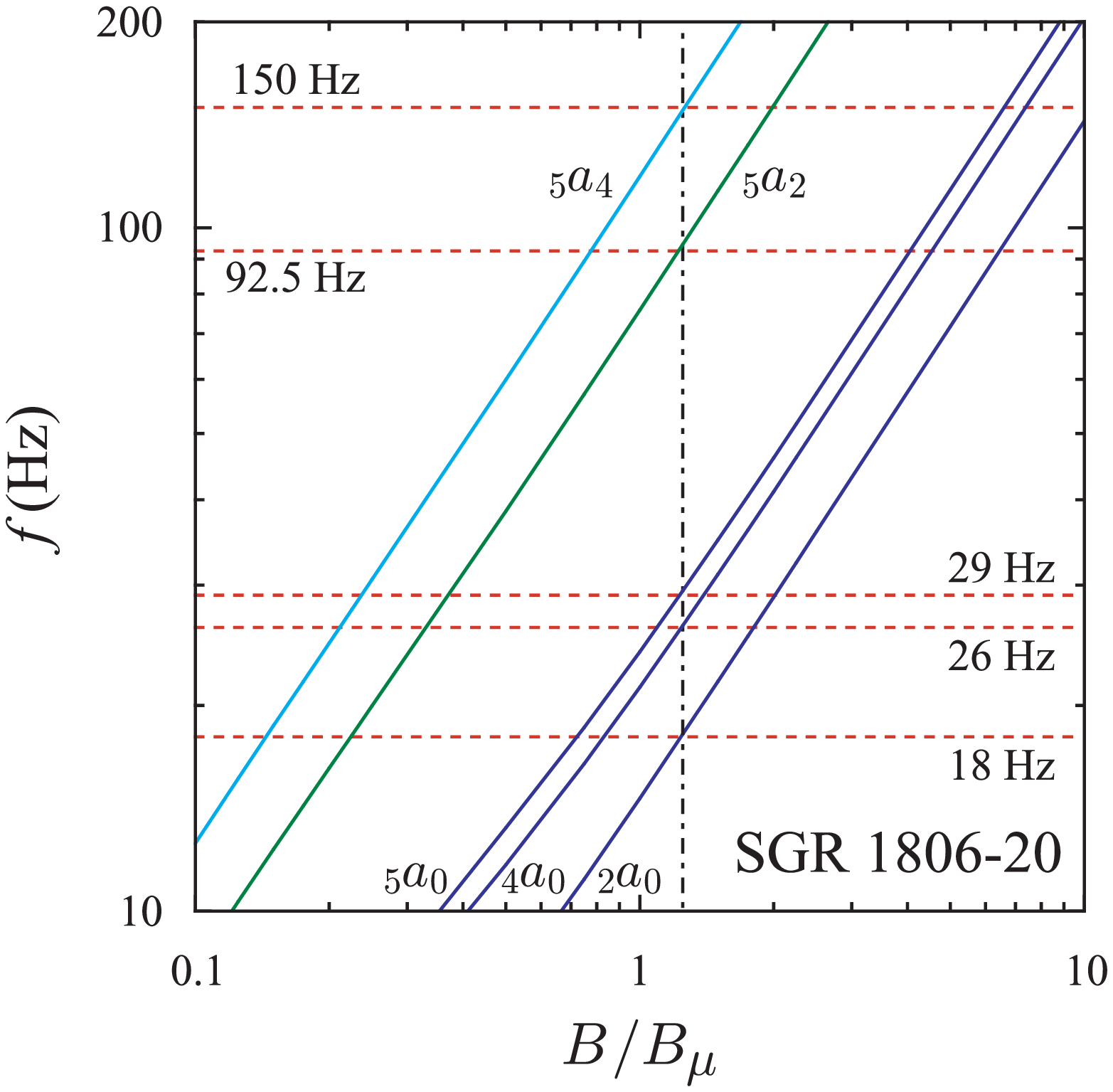}
%\vspace{5mm}
 \caption{Comparison of several specific Alfv\'{e}n modes (solid lines) 
   with observed frequencies for SGR 1806-20 (horizontal dashed lines)
   for the stellar model A+DH$_{14}$. The agreement for all
   frequencies up to 150 Hz is excellent, for a magnetic field
   strength of $B/B_{\mu} \approx 1.25$ (vertical
   dashed-dotted line).}
  \label{Fig:ADH14-1806}
\end{figure}

Figure \ref{Fig:ADH14-1806} shows, as an example, various global
Alfv\'{e}n mode frequencies for model A+DH$_{14}$.  In this figure,
the horizontal broken lines denote the observational data up to 150 Hz
for SGR 1806-20 while the solid lines denote our computed frequencies.
We find an excellent agreement between observed and computed
frequencies for a magnetic field strength of $B/B_{\mu}\approx 1.25$,
(shown as a vertical dash-dotted line). Specifically, the observed
frequencies of of 18, 26, 29, 92.5, and 150 Hz can be identified with
the ${}_2a_0$, ${}_4a_0$, ${}_5a_0$, ${}_5a_2$, and ${}_5a_4$ modes,
respectively.

\begin{figure}
\includegraphics[width=80mm]{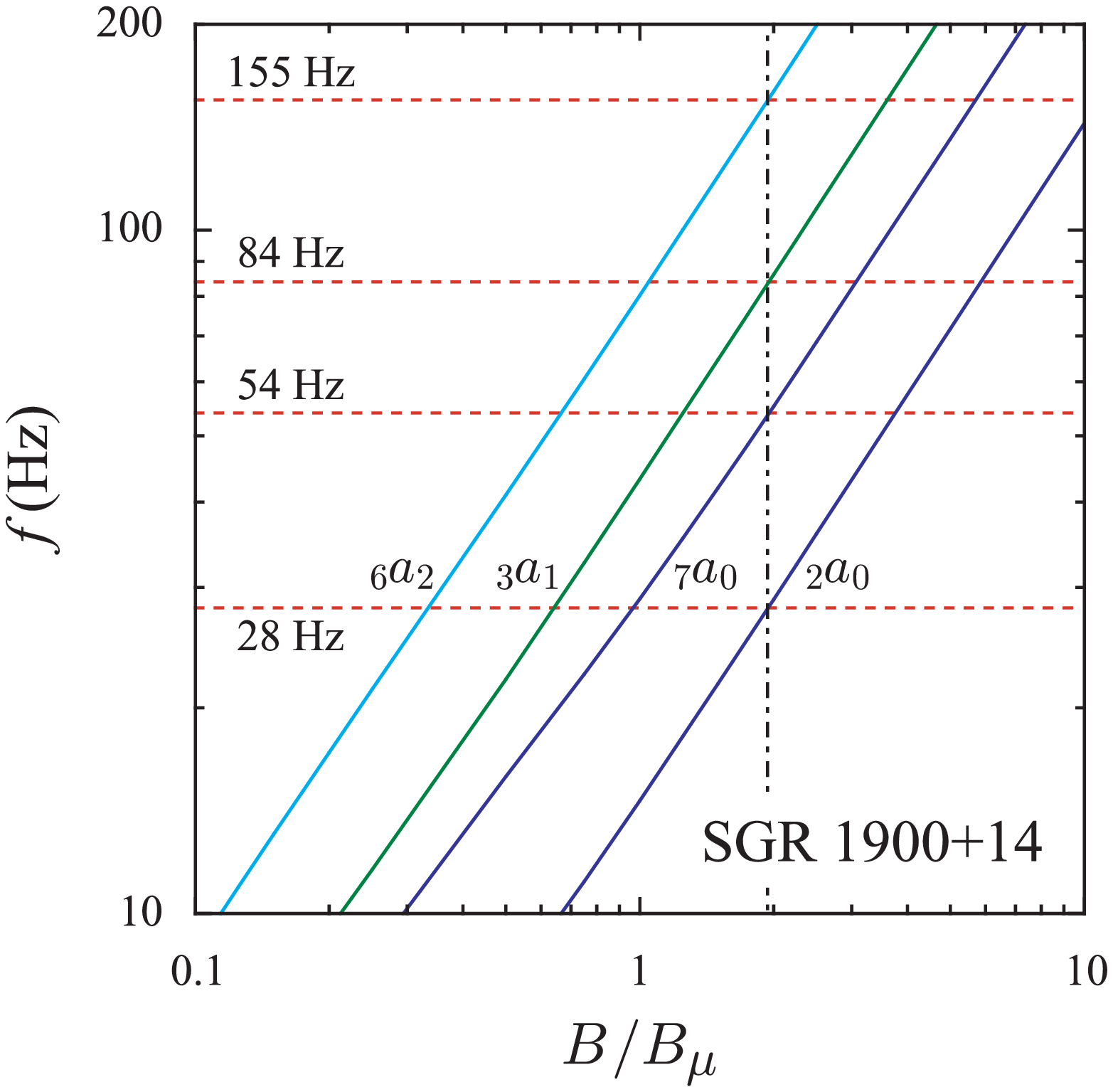}
%\vspace{5mm}
 \caption{Comparison of several specific Alfv\'{e}n modes (solid lines) 
   with observed frequencies for SGR 1900+14 (horizontal dashed lines)
   for the stellar model A+DH$_{14}$. The agreement for all
   frequencies up to 150 Hz is excellent, for a magnetic field
   strength of $B/B_{\mu} \approx 1.94$ (vertical
   dashed-dotted line).}
  \label{Fig:ADH14-1900}
\end{figure}

Figure \ref{Fig:ADH14-1900} shows a similar excellent agreement of the
observed frequencies for SGR 1900+14, with the frequencies of the
${}_2a_0$, ${}_7a_0$, ${}_3a_1$, and ${}_6a_2$, respectively, for the
same chosen stellar model A+DH$_{14}$ and a slightly larger magnetic
field strength. Since the global frequencies
are sensitive to both the stellar parameters and the magnetic
field strength, an independent determination of the compactness of the
star would allow for a determination of the magnetic field.

The higher-frequency observational data for SGR 1806-20, such as
626.5, 720, 1837, and 2384 Hz could be also be identified with higher
overtones of global Alfv\'{e}n modes, as there exist many higher
overtones with different $\ell$.  For example, for model A+DH$_{14}$
and with $B/B_{\mu} = 1.25$ (as above), we compute the following mode
frequencies: ${}_4a_{25}=632$, ${}_4a_{29}=724$, ${}_2a_{83}=1841$,
and ${}_2a_{108}=2382$ Hz, which are in all good agreement with the
observed frequencies.  We remark that both the  626.5 and 720 Hz
frequencies can be identified with some global Alfv\'{e}n modes, while
this would not be possible with the crustal modes, when the magnetic
field is confined to the crust (see numerical results in Paper I),
due to the near degeneracy of overtones of different $\ell$.

Table \ref{Tab:fit_magnetic-field} shows the maximum values of the
magnetic field, for which one can at least identify the frequency of
the ${}_2a_0$ mode. For models which have no entry in this Table, such
an identification was not possible.  In reality, some stellar models
with lower magnetic strength can agree well with the observational
data, if the ${}_2a_0$ mode is omitted.  For example, the stellar
model with L+NV$_{18}$ and $B/B_{\mu}=0.60$ can agree well with all
observational data lower than 150 Hz, when identified with the
frequencies of the ${}_4a_0$, ${}_8a_0$, ${}_{10}a_0$, ${}_2a_6$, and
${}_2a_{11}$ modes respectively.  As seen Figure
\ref{Fig:A_DH_n01_l234} there are no global Alfv\'{e}n modes with
frequency less than that of the ${}_2a_0$ mode. Thus, the values in
Table \ref{Tab:fit_magnetic-field} correspond to the maximum magnetic
field strength for each stellar model, if the observational data could
be explained by the global Alfv\'{e}n modes. The maximum magnetic field
strength is $B/B_{\mu}\approx 2$ for SGR 1806-20 and $B/B_{\mu}\approx
3$ for SGR 1900+14.

\begin{table*}
 \centering
 \begin{minipage}{80mm}
  \caption{Maximum magnetic field strength, for various equilibrium models,
    for which the lowest frequency observed in two different SGR sources can be
    identified with the fundamental ${}_{2}a_0$ mode.}
\label{Tab:fit_magnetic-field}
  \begin{tabular}{@{}lcc@{}}
  \hline
   Model  & SGR1806-20 & SGR1900+14  \\
 \hline
 A+DH$_{14}$    & 1.25     & 1.94      \\
 A+DH$_{16}$    & 1.65     & 2.64      \\
 WFF3+DH$_{14}$ & 1.04     & 1.65      \\
 WFF3+DH$_{16}$ & 1.26     & 2.00      \\
 WFF3+DH$_{18}$ & 1.67     & 2.68      \\
 APR+DH$_{14}$  & 0.89     & 1.43      \\
 APR+DH$_{16}$  & 1.07     & 1.69      \\
 APR+DH$_{18}$  & 1.28     & 2.00      \\
 APR+DH$_{20}$  & 1.52     & 2.36      \\
 APR+DH$_{22}$  & 1.84     & 2.93      \\
 L+DH$_{14}$    & ---      & 1.20      \\
 L+DH$_{16}$    & 0.87     & 1.38      \\
 L+DH$_{18}$    & 1.00     & 1.58      \\
 L+DH$_{20}$    & 1.15     & 1.82      \\
 L+DH$_{22}$    & 1.32     & 2.05      \\
 L+DH$_{24}$    & 1.52     & 2.39      \\
 L+DH$_{26}$    & 1.80     & 2.85      \\
\hline
 A+NV$_{14}$    & ---      & 1.92      \\
 A+NV$_{16}$    & 1.63     & ---       \\
 WFF3+NV$_{14}$ & ---      & 1.58      \\
 WFF3+NV$_{16}$ & 1.21     & 2.00      \\
 WFF3+NV$_{18}$ & 1.65     & ---       \\
 APR+NV$_{14}$  & ---      & ---       \\
 APR+NV$_{16}$  & ---      & 1.70      \\
 APR+NV$_{18}$  & 1.25     & 2.00      \\
 APR+NV$_{20}$  & 1.51     & 2.42      \\
 APR+NV$_{22}$  & 1.85     & 2.98      \\
 L+NV$_{14}$    & ---      & ---       \\
 L+NV$_{16}$    & ---      & ---       \\
 L+NV$_{18}$    & ---      & 1.69      \\
 L+NV$_{20}$    & ---      & 1.94      \\
 L+NV$_{22}$    & 1.39     & 2.24      \\
 L+NV$_{24}$    & 1.63     & 2.58      \\
 L+NV$_{26}$    & 1.92     & ---       \\
\hline
\end{tabular}
\end{minipage}
\end{table*}

%%%%%%%%%%%%%%%%%%%%%%%%%%%%%%%%%%%%%%%%%%%%%%%%%%%%%%%%%%%%%%%%%%%%%%%%%%%%
%%%%%%%%%%%%%%%%%%%%%%%%%%%%%%%%%%%%%%%%%%%%%%%%%%%%%%%%%%%%%%%%%%%%%%%%%%%%
\section{Summary and Discussion}
\label{sec:V}
%%%%%%%%%%%%%%%%%%%%%%%%%%%%%%%%%%%%%%%%%%%%%%%%%%%%%%%%%%%%%%%%%%%%%%%%%%%%
%%%%%%%%%%%%%%%%%%%%%%%%%%%%%%%%%%%%%%%%%%%%%%%%%%%%%%%%%%%%%%%%%%%%%%%%%%%%

We have investigated torsional Alfv\'{e}n modes of relativistic stars
with a global dipole magnetic field. It has been noted recently
(Glampedakis et al.  2006) that such oscillation modes could serve as
as an alternative explanation (in contrast to torsional crustal modes)
for the SGR phenomenon, if the magnetic field is not confined to the
crust. We compute global Alfv\'{e}n modes for a representative sample
of equations of state and magnetar masses, in the ideal MHD
approximation and ignoring $\ell \pm 2$ terms in the eigenfunction.
The presence of a realistic crust has a negligible effect on
Alfv\'{e}n modes for $B > 4\times 10^{15}$ G.  Using a toy model with
large crust thickness we demonstrate that strong avoided crossings
between torsional Alfv\'{e}n modes and torsional crust modes take
place. For magnetar-like magnetic field strengths, the spacing between
consecutive Alfv\'{e}n modes is of the same order as the gap of
avoided crossings. As a result, it is not possible to identify modes
of predominantly crustal character and all oscillations are
predominantly Alfv\'{e}n-like.

The Alfv\'{e}n-like character of global modes allows for a
non-degeneracy of overtones with different $\ell$. Using this fact, we
find excellent agreement between our computed frequencies and observed
frequencies in two SGRs, for reasonable magnetic field strengths. We
construct an empirical formula which describes the frequency of the
fundamental torsional Alfv\'{e}n modes (for different $\ell$) as a
function of the magnetic field strength and the compactness of the
star. Such formulae could be used for inferring the stellar properties
and magnetic field strength. The requirement that the lowest observed
frequency is identified with the fundamental ${}_2a_0$ mode leads to a
maximum magnetic field strength in the range of (0.8--1.2)$\times
10^{16}$ G.

In the present work we have ignored couplings to $\ell \pm 2$ terms.
A two-dimensional treatment of the eigenvalue problem will appear
elsewhere.

%\newpage
%%%%%%%%%%%%%%%%%%%%%%%%%%%%%%%%%%%%%%%%%%%%%%%%%%%%%%%%%%%%%%%%%%%%%%
\section*{Acknowledgments}
%%%%%%%%%%%%%%%%%%%%%%%%%%%%%%%%%%%%%%%%%%%%%%%%%%%%%%%%%%%%%%%%%%%%%%

%We are grateful to Nils Andersson and Adamantios Stavridis for useful discussions.
This work was supported by the Marie-Curie grant
MIF1-CT-2005-021979, the Pythagoras II program of GSRT and the EU
network ILIAS.

%%%%%%%%%%%%%%%%%%%%%%%%%%%%%%%%%%%%%%%%%%%%%%%%%%%%%%%%%%%%%%%%%%%%%%%%%%%%
%%%%%%%%%%%%%%%%%%%%%%%%%%%%%%%%%%%%%%%%%%%%%%%%%%%%%%%%%%%%%%%%%%%%%%%%%%%%
%\begin{thebibliography}{999}

%\appendix
%%%%%%%%%%%%%%%%%%%%%%%%%%%%%%%%%%%%%%%%%%%%%%%%%%%%%%%%%%%%%%%%%%%%%%%%%%%%
%%%%%%%%%%%%%%%%%%%%%%%%%%%%%%%%%%%%%%%%%%%%%%%%%%%%%%%%%%%%%%%%%%%%%%%%%%%%
%\section[]{}
%\label{sec:A1}
%%%%%%%%%%%%%%%%%%%%%%%%%%%%%%%%%%%%%%%%%%%%%%%%%%%%%%%%%%%%%%%%%%%%%%%%%%%%
%%%%%%%%%%%%%%%%%%%%%%%%%%%%%%%%%%%%%%%%%%%%%%%%%%%%%%%%%%%%%%%%%%%%%%%%%%%%

\end{document}